\documentclass[pre,aps,twocolumn,floatfix]{revtex4}
\usepackage{graphicx}
\usepackage{epsfig}

\begin{document}

\date{\today}
\title{A short-loop algorithm for quantum Monte Carlo simulations}

\author{Ying-Jer Kao}
\email{yjkao@phys.ntu.edu.tw}
\affiliation{Department of Physics and Center for Theoretical Sciences, National Taiwan University, Taipei, Taiwan 106}
\date{\today}

\author{Roger G. Melko}
\email{rgmelko@science.uwaterloo.ca}
\affiliation{Department of Physics and Astronomy, University of Waterloo, Ontario, N2L 3G1, Canada}
\affiliation{Materials Science and Technology Division, Oak Ridge National Laboratory, Oak Ridge TN, 37831, USA}

\begin{abstract}
We present an algorithmic framework for a variant of the quantum Monte Carlo operator-loop algorithm, where non-local cluster updates are constructed 
in a way that makes each individual loop smaller.  The algorithm is designed to increase simulation efficiency in cases where conventional loops become very large, do not close altogether, or otherwise behave poorly.  We demonstrate and characterize some aspects of the short-loop on a square lattice spin-1/2 XXZ model where, remarkably, a significant increase in simulation efficiency is observed in some parameter regimes.  The simplicity of the model provides a prototype for the use of short-loops on more complicated quantum systems.
\end{abstract}
\maketitle

\section{introduction}

Quantum Monte Carlo (QMC) simulations \cite{Suz,hands1} comprise arguably the most powerful set of methods for analyzing
strongly-interacting models in quantum many-body physics.  Breakthroughs in QMC methodology 
over the last decade or so have enabled the study of simulation cells of unsurpassed finite size,  
many capable of simulating millions of
quantum species
 for simple models.  Traditionally, large system sizes were
coveted to enable finite-size scaling to the thermodynamic limit, something that remains 
important for the study quantum ground states and critical phenomena, where unconventional or non-monotonic scaling is sometimes observed \cite{JKprl}.  However, recent
interest in nanoscale quantum systems, as well as ultra-cold atoms trapped in optical lattices, has
provided a situation where QMC methods are able to approach realistic experimental systems sizes
\cite{optical}.  The work on algorithmic advances therefore continues at a rapid pace.

Besides the infamous sign problem \cite{Sign,troyersign}, which precludes the simulation of many fermionic and frustrated 
magnetic systems, the largest general obstacle for QMC methods are algorithm freezing, critical slowing
down, or other phenomena perhaps best summarized as ``loss of ergodicity''.  These can result in 
problems ranging from a slight loss of efficiency (requiring longer Monte Carlo runs to reach
a desired level of statistical accuracy), to serious issues such as complete non-ergodicity in some
parameter regimes, leading to the obscuration of all interesting physics in the model.  For example, an inability to
accurately measure a subset of estimators (in particular off-diagonal quantities) is a drawback
of some classes of simple  ``local'' QMC updates \cite{KawHarRev}.

Perhaps the most important algorithmic breakthrough in QMC technology was the introduction
of the {\it loop} algorithm by Evertz, Lana and Marcu \cite{Evertz}.  Until that time, the QMC sampling
procedure proceeded via local updates, roughly analogous to single spin flips in a simple Monte Carlo
simulation of a classical Ising model \cite{KawHarRev}.  The loop algorithm, analogous to a Wolff or 
Swenson-Wang cluster (or global) update, solved ergodicity problems related to sampling in a 
grand-canonical framework, and also facilitated the measurement of some off-diagonal quantities.
Originally formulated in a discrete world-line framework, the algorithm has been continually refined 
and advanced, and is widely used in all modern QMC frameworks, for example in continuous world-line 
methods (including worm algorithm variants \cite{PSTworm,Bonn_Worm}) and the stochastic series expansion (SSE) framework, which employs the  ``operator'' \cite{sse4} or  ``directed''-loop variants \cite{DIRloop}.

The common feature of all QMC loop algorithms is the creation of a defect or singular point (or in the 
case of the worm algorithm, two points) which propagates through the simulation cell updating the QMC
representation of the Hamiltonian or partition function (i.e.~the world-line configuration, or the basis
state and operator-list).  This defect is typically resolved when it encounters its starting point, (or 
 another propagating defect) forming a single closed loop.  Loops formed in this way may then be used in
a variety of single- or multi-cluster sampling schemes \cite{KawHarRev,LiuMC}.
In the following, we will call such an algorithm, where the closing condition of the loop is 
that its ``head'' meets its ``tail'', a conventional or {\em{long-loop}}.

In classical Monte Carlo methods, the prototypical analogy of the above algorithm was first introduced for 
the problem of proton distribution in ice water \cite{Rahman}, and later extended to Monte Carlo simulations
of other vertex and ice models \cite{BarkNew}.  
This classical loop algorithm effectively allows targeted updates in a reduced manifold of low-energy 
vertex states.  The original classical loop is the long-loop, as described above (see also Ref.~\cite{ClassicalWorm}), however a variation
that involves loops of shorter length has been shown to perform more efficiently in a large number of
cases, and has become widely adopted \cite{BarkNew,RogMichRev}.  
This variation became known as the {\em short-loop} algorithm, and as its name implies, involves
creating loops of much smaller total length.
A key reason for the increase in efficiency observed with short-loops appears to be a respite from
the tendency of long-loops to grow in proportion to the size of the simulation cell,
which in some cases can result in excessively long updates and a delay in defect resolution \cite{RogMichRev}.  Additionally, short loops do not have the capacity to re-trace multiple paths through the same region of configuration space, avoiding the wasted computational overhead that often can occur in long-loop algorithms. 

\begin{figure}[floatfix] {
\includegraphics[width=2.4 in]{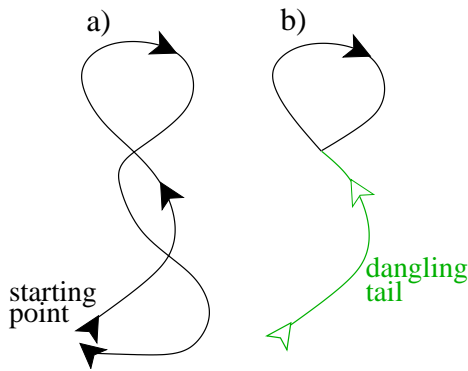}
\caption{(color online) 
Schematic comparison of a long-loop (a) versus a short-loop (b).  In (a), the loop defect propagates until it
encounters its own starting point.  In (b), the loop defect propagates only until it encounters its own path.  The 
dangling tail (green line) must be removed.
\label{SLloop_1}}
}
\end{figure}

Conceptually, such {\em short-loops} are distinguished from the long-loop construct based simply on the closing 
or resolution condition of the loop's head or defect.
Namely, a short-loop closes not only if the defect encounters its own starting point (ie.~the 
head meets the tail), but also if it encounters any other previous point of the loop body.  Short-loops 
are also differentiated by the resulting {\it dangling tail} of propagated defects, which must be removed from the 
loop structure before the Monte Carlo update can continue (see Fig.~\ref{SLloop_1}).

From this description , the classical definition \cite{BarkNew,RogMichRev} of the 
short-loop algorithm can be adapted to the case of an operator-loop algorithm in a $d+1$ quantum simulation cell.
In this paper, we provide a detailed description of the short-loop algorithm in a full QMC framework.
We note that short-loops may be 
formulated in any of the aforementioned QMC algorithmic flavors; in the next section we choose to 
introduce them in the popular and simple SSE QMC paradigm \cite{sse4,sse1,sse2,DIRloop}.
We are particularly motivated by the question of whether the large efficiency gains enjoyed by short-loops
in classical Monte Carlo simulations of vertex models will translate over to the QMC arena.
In Section~\ref{autoC}, we attempt to answer this question with concrete autocorrelation measurements 
on the simple demonstrative case of the two-dimensional (2D) $S=1/2$ XXZ model. We conclude the paper 
with a short discussion of several advantages and disadvantages of the short-loop algorithm,
and possible adaptations of it to more complicated quantum models in the future.

\section{Loop algorithms in the stochastic series expansion framework}

The SSE decomposition of a quantum Hamiltonian on a $d$-dimensional lattice proceeds via the 
expansion of the finite-temperature partition function \cite{sse1,DIRloop},
\begin{equation}
Z = \sum\limits_\alpha \sum_{n=0}^{\infty}  \sum_{S_n} \frac{(- \beta)^n}{n!}
    \left \langle \alpha  \left | \prod_{i=1}^n H_{t_i,b_i}
    \right | \alpha \right \rangle .
\label{Zpart}
\end{equation}
Here, the sum over $S_n$ represents a sampling of an operator-index sequence (defined below),
performed via a Metropolis Monte Carlo procedure.
In $Z$, a quantum Hamiltonian is typically written as a sum of elementary interactions,
\begin{equation}
H = - \sum\limits_{t}\sum\limits_{b} H_{t,b},
\label{hsum}
\end{equation}
where in a chosen basis $\{ |\alpha \rangle \}$ (e.g.~the standard $S^z$ basis) the operators satisfy
$H_{t,b}|\alpha \rangle \sim |\alpha^\prime \rangle$,
and $|\alpha \rangle$ and $|\alpha^\prime \rangle$ are both basis states.
The index $t$ refers to the operator types (various kinetic and
potential terms), while $b$ is the lattice unit over which the interactions are
summed (e.g.~a nearest-neighbor bond) .  The operator-index sequence is hence represented as 
$S_n = [t_1,b_1] \ldots [t_n, b_n]$, where $n$ is the expansion order.
Typically, the size of the operator-index sequence is set to some constant $M>n$ (since $n$ fluctuates), and the
operator-index list is filled in with unit or identity operators, represented in $S_M$ as $[0,0]$.

For concreteness, we will consider the paradigmatic spin-1/2 XXZ model, 
\begin{equation}
H = J \sum_{\langle i,j \rangle} ( S^x_iS^x_j + S^y_iS^y_j +\Delta S^z_i S^z_j)  - h\sum_i S^z_i.
\label{XXZham}
\end{equation}
A standard SSE algorithm for this Hamiltonian is laid out in detail in Ref.~\cite{DIRloop}, and we 
refer the reader to that work as we make frequent reference to it in the upcoming discussion.
In particular, the square lattice decomposition (Eq.~(\ref{hsum})) for this Hamiltonian results in two bond terms,
\begin{eqnarray}
H_{1,b}/J &=& C - \Delta S^z_i S^z_j + \frac{h}{4J}(S^z_i +S^z_j), \\
H_{2,b}/J &=& \frac{1}{2}(S^+_i S^-_j + S^-_i S^+_j ),
\end{eqnarray}
where the constant $C$ is defined as necessary to make $H_{1,b}>0$, hence avoiding the sign problem.

\begin{figure}{
\includegraphics[width=1.4 in]{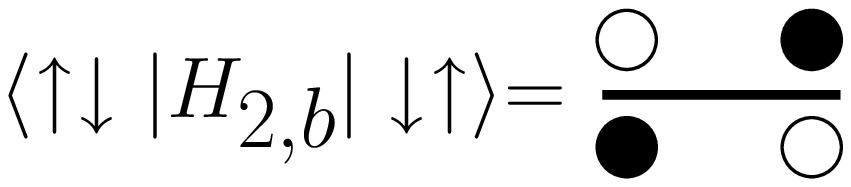}
\caption{(color online) 
A vertex as a graphical representation of a bond matrix element.  Filled circles represent spin $+1/2$, open
represent spin $-1/2$.  For the XXZ model (Eq.~(\ref{XXZham})) this vertex has weight $J/2$ \cite{DIRloop}.
\label{Vertex}}
}
\end{figure}

There are two standard (non-trivial) updates for SSE simulations of typical Hamiltonians.  The first
is the {\it diagonal update}, designed to perform substitutions $[0,0] \leftrightarrow [1,b]$, changing
the expansion order $n$.
The second update, of interest to us, is the {\it operator loop} update, which accomplishes 
substitutions within and between operator-list elements $[1,b]$ and $[2,b]$, keeping $n$ fixed but 
effectively sampling off-diagonal
operators.  The operator loop is performed in a linked list of {\it vertices}, an abstraction of the propagation of the basis state $| \alpha \rangle$ by $S_M$ in the $d+1$ dimensional simulation cell \cite{DIRloop}.  The linked list is
defined graphically
by single operators propagating a unit's (bond's) basis state at some given expansion step (see Fig.~\ref{Vertex}). 
In the $S=1/2$ XXZ model, there are six allowed vertices resulting from six non-zero matrix elements
(see Eq.~(18) of Ref.~\cite{DIRloop}).

In the conventional long-loop SSE algorithm, a vertex is updated by a propagating defect.
The defect propagates along the linked list and, upon meeting its own starting point (ie.~when the head meets the tail), forms a closed loop. 
Typically, the starting point of the loop is chosen randomly from the linked vertex list.
During the propagation, the defect enters a vertex simply by following a link from the ``exit-leg'' of
the previously visited vertex.  An exit-leg is typically chosen by a Metropolis Monte Carlo procedure: for example,  a simple heat-bath scheme where the probability of exiting along any given vertex leg is proportional to the weight of the resulting matrix element.  
A particularly efficient way to choose these exit probabilities in the SSE is to use the directed-loop
equations, detailed in Ref.~\cite{DIRloop} - however the form of the loop algorithm (long or short) is independent of the choice of exit probabilities. 

Once closed, a long loop satisfies detailed balance and, in effect, the visited
vertex legs may be flipped with 
probability 1 -- hence its relationship to the classical Wolff cluster algorithm \cite{LiuMC}.  In practice, one
need not store the loop path at all, as updating of the vertex legs occurs in real-time as the 
defect propagates.  Once closed, the vertices visited by the loop are already affected (ie. flipped)
and one must simply update the stored global basis state $| \alpha \rangle$ and operator-index sequence $S_M$.
Note that this update typically occurs after a significant number of loops have been preformed - this number is discussed much more below.  

As alluded to above, one difficulty encountered in this loop algorithm for some parameter regimes of
certain Hamiltonians (not necessarily Eq.~(\ref{XXZham})) is that loops can become very long before
they close, or sometimes in extreme cases do not close at all \cite{AndersPC,PTsse}.  The standard practice to
combat this is to impose some maximum loop length
\begin{equation}
{\ell}^{\rm max}_{\rm loop} = c_o n,
\label{Llength}
\end{equation}
($c_o$ is some constant),  upon reaching which loop construction is terminated.  Here, loop length
may be measured for example in the number of vertex legs traversed per algorithm iteration (typically two).
In the case of termination, detailed balance is preserved by disregarding updates attained by the loop, and keeping
the previous Monte Carlo step's $| \alpha \rangle$ and $S_M$ \cite{DIRloop}.
Unfortunately, the algorithm overhead (i.e.~CPU time) used in constructing the aborted loop or set of loops
is lost in this case.

These examples serve to further motivate the development of a loop algorithm that does not suffer from such
drawbacks.  One solution, in analogy to the classical short-loop algorithms discussed in the previous
section, is a quantum short-loop variant of the conventional SSE operator loop.  
In the next section, we discuss details of the quantum short-loop algorithm, including the closing condition,
handling of bounce processes, and tail removal.  

\section{The short-loop algorithm}

\subsection{Overview}
\label{SLov}

At first glance, the definition of the short-loop algorithm in the SSE is quite simple.  Begin by propagating
a loop defect as one would normally do for the long operator-loop, starting from a random vertex-leg.  
In the event where the propagating defect 
encounters a vertex-leg where it has previously been, terminate the loop algorithm.  The segment of the path created by the defect that does not form the loop is the {\em dangling tail} (see Fig.~ \ref{SLloop_1}), and must be removed or reverted back to its original state.  Also, a consequence of the need to remove this tail is the requirement to {\it store} the loop path created by the propagating defect - something that is not needed in the conventional long-loop algorithm. 

\begin{figure}[floatfix] {
\includegraphics[width=3.3 in]{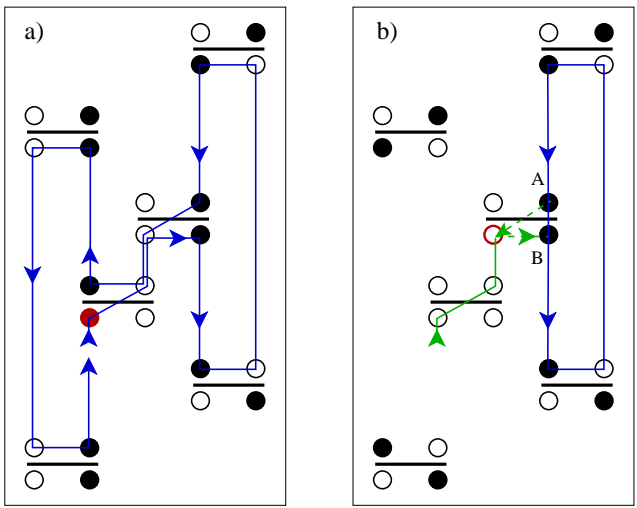}
\caption{(color online) 
A linked list of six vertices, with operator loops and {\it final} (ie. flipped) basis state spins.
For clarity, links are not illustrated, but occur as vertical lines connecting vertex legs \cite{DIRloop}.
(a) A long-loop, which propagates around the linked list until its encounters its own original
starting vertex (red).
(b) The same loop construction, if governed by short-loop rules, would propagate around the linked 
list until it encounters a previously-visited vertex leg which was also an {\it in}-leg (red).
The loop closes in the vertex containing this leg, by connecting the remaining visited sites (A and B).  The beginning portion of the loop propagation, the {\it dangling tail} (solid green line), is removed and the
associated vertex legs are not flipped.
\label{SLloop1}}
}
\end{figure}

Consider the important closing condition of the short-loop algorithm in more detail.
It turns out that, unlike the classical case, the simplified criterion mentioned above (the loop closes upon encountering any previously-visited vertex leg) is insufficient for the QMC case, since a quantum operator vertex is involved. 
To facilitate closing of the quantum short-loop, the terminating leg should have been,
upon its original visit, an {\em in}-leg (see Fig.~\ref{SLloop1}); if the propagating defect encounters
a previously-visited {\it out}-leg, the loop creation algorithm should continue unabated.
An attempt to close the short-loop at an \textit{out}-leg
would result in an un-resolvable defect, where removing the dangling tail becomes impossible without destroying the loop itself.  
Once the terminating leg is chosen, its spin is not flipped, and the loop is closed at that vertex using the remaining two visited legs, finally resolving the propagating defect.
Assuming that one has flipped spins associated with vertex legs during the propagation of the defect,
starting from the terminating vertex leg,  flip back all spins and update the associated vertices on the dangling tail, until the initial starting point is reached.  
Since this tail removal occurs deterministically, detailed balance remains satisfied for the short-loop algorithm.   The short-loop is now complete, and the usual progression of the 
SSE Monte Carlo algorithm (i.e.~more loop updates, diagonal updates or measurements) may proceed.

A comparison between two long- and short-loops is illustrated in Fig.~\ref{SLloop1}.  Even in this simple 
case, several key factors that determine loop efficiency are apparent.  First, the short-loop is obviously much
smaller (in the total number of vertices visited) than the long-loop, which served as original motivation
for designing this algorithm. 
Further, by
inspecting the center of Fig.~\ref{SLloop1}a, it is apparent that two vertex-legs have been visited twice
in this illustration.  Processes such as this (``re-tracing'') may have a negative impact on long-loop 
performance, since the computational effort associated with propagating the defect through these vertex-legs may ultimately result in no flipped basis spins.  In this way, it is apparent that a long-loop could correspond to constructing separate smaller loops and flipping them all {\it together}.  In contrast, the same smaller loops, if constructed with several short-loop updates, would flip the loops {\it independently}.  Issues like this may also result in slight improvements in efficiency when using the short-loop algorithm.

Any gains of this sort must however be weighed against the computational overhead associated with storing the short-loop, resolving the propagating defect in the terminating vertex, and removing the dangling tail.  The process is illustrated in Fig.~\ref{SLloop1}b, where the 
dangling tail involves three vertex-legs (two of which must be re-flipped).
In addition
to this computational overhead, additional data structures are required in the formation of the short-loop
algorithm to store the loop path in memory to allow for the removal of the tail.  
Since the additional CPU time and memory burdens may conceivably negate any efficiency improvements gained on the long-loop, the simple arguments associated with Fig.~\ref{SLloop1} are
likely not sufficient to draw quantitative conclusions of short-loop performance -- this is left to 
Section~IV where we discuss autocorrelation results.  Before addressing this, we proceed with several more key details to note when implementing the short-loop algorithm in a practical QMC code.
                                                                               
\subsection{Short-loops in the presence of bounce processes}

In the previous discussion, one important complicating factor was purposely neglected: the handling of so-called
{\it bounce} processes in the formation of the operator loops.  Bounce processes (see also Ref.~\cite{DIRloop}) 
are defined as the case where a propagating loop defect, upon encountering any given vertex, chooses (by way
of the specific Hamiltonian and algorithmic probability tables) its {\it out}-leg to be the same as its
{\it in}-leg, thereby starting on a path which re-traces, for some distance at least, the loop back along its
path of previously-visited vertex legs.

Bounce processes are known to be the most serious detriment to the efficiency of the loop update in the 
SSE \cite{DIRloop} (although they are perhaps not the only detrimental process \cite{GEN_dirloop}).  Advanced methods
to construct the QMC probability tables governing loop propagation, such as the directed-loop 
weights \cite{DIRloop,GEN_dirloop,pollet}, combat this problem by minimizing the weight of bounce processes when possible.  However, it is 
common to find many physically interesting models where bounce processes cannot be avoided.  As such, any
practical implementation of a short-loop algorithm must be able to take bounces into account.

\begin{figure}[floatfix] {
\includegraphics[width=1.6 in]{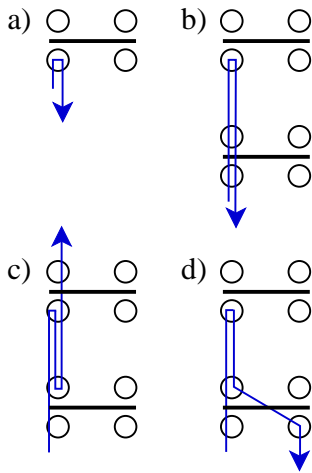}
\caption{(color online) 
(a) A bounce process which: (b) continues to re-trace the loop path; (c) re-bounces to continue a new path; 
(d) branches to continue a new path.
\label{BounceP}}
}
\end{figure}
The simple short-loop algorithm described in Section~\ref{SLov} requires several modifications, which are essentially
conditions to ensure that the loop doesn't terminate prematurely if it encounters its own path due to a bounce process.   Recall that, in order to remove the dangling tail upon termination, the short-loop requires
storage of the loop propagation-path.  Considering the possibility of bounce processes, it becomes obvious that a 
modified {\it stack} (last-in, first-out) is the appropriate data structure with which to store the loop propagation.
Vertex-legs which are part of a new path should be pushed onto the top of the stack, while bounces
or re-traced vertex-legs should be popped off the top.
More specifically, several cases should be considered (see Fig.~\ref{BounceP}).  First, in the
simple case of a first bounce, easily identifiable since the in-leg is identical to the 
out-leg, the first bounce vertex-leg is not added to the stack (Fig.~\ref{BounceP}a).
If the bounce continues along
the previously-visited path (Fig.~\ref{BounceP}b), previously-visited vertex legs are popped off the top
of the stack.  New legs are added to the stack when the path deviates from the previously-visited path,
as in cases Fig.~\ref{BounceP}c and Fig.~\ref{BounceP}d.  Note that only these last two cases, 
where the propagating defect begins tracing a new path, should have the option of closing the short-loop.

After the short-loop eventually closes, this stack data structure is accessed from the 
terminating vertex-leg and re-traced to the {\it bottom} of the stack to remove the dangling tail (see Fig.~\ref{SLloop1}).  With bounce processes pushed and popped correctly during defect propagation,
the algorithmic overhead involved with removing the tail can be reduced considerably.

With these considerations, the problem of implementing the short-loop algorithm is essentially a 
coding procedure -- efficient execution of storage and propagation processes.  CPU and memory 
requirements can vary considerably depending on this implementation.  In the section below, we provide
some quantification of the short-loop efficiency with one particular C++ implementation, using
 standard STL data structures.

\section{Simulation results for the XXZ model}
\label{autoC}

\begin{figure}[floatfix] 
\includegraphics[width=3.0 in]{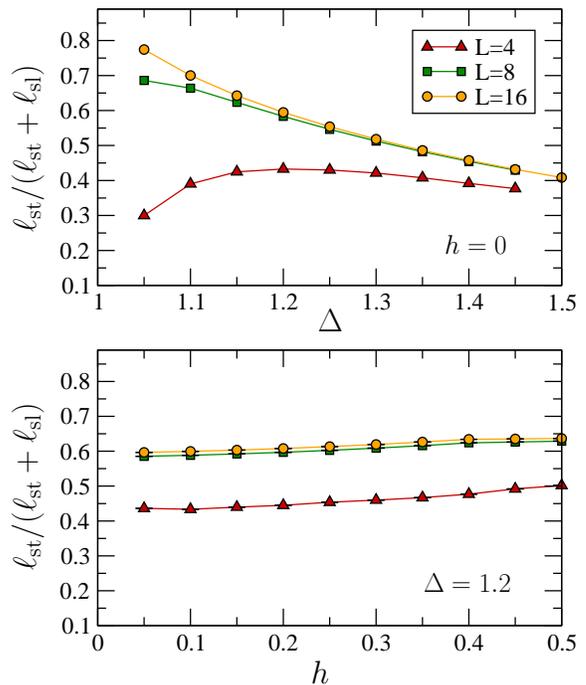}
\caption{(color online) 
Tail-length as a fraction of the total cluster size, measured in number of vertices visited, for simulations of the 2D XXZ model employing the short-loop algorithm
and one particular solution of the directed-loop equations \cite{DIRloop}.  Here, $\beta = 2L$.
\label{DHtail}}
\end{figure}

For concreteness, we present a comparison of the efficiency of the long- and short-loop algorithms in the 
simple XXZ model, Eq.~(\ref{XXZham}), where in the below data we have set $J=1$.
One of the first important indicators of short-loop efficiency is the length of the dangling tail, i.e.~the
discarded list of vertices that are not included in the definition of the closed loop for the purposes of
updating the simulation.  Long tails are generally detrimental to loop efficiency, due to the wasted CPU 
effort in both constructing and erasing them.  In Fig.~\ref{DHtail}, this ratio is illustrated for several simulation sizes and parameter values, where the ratio of the tail length ($\ell_{\rm st}$) versus the 
total cluster size (tail plus loop: $\ell_{\rm st} +\ell_{\rm sl}$) is plotted.
Here, the short-loop length ($\ell_{\rm sl}$) is defined without bounce or back-tracking processes 
(see Fig.~\ref{BounceP}).
From this figure, it is clear that the ratio $\ell_{\rm st}/\ell_{\rm sl}$ depends highly on simulation parameters,
however one tends to see convergence in parameter regions of large $\Delta$ or $h$, particularly
with system size.
This demonstration shows that the tail in the short loop algorithm is, for these parameters, on the order of
the length of the loop itself.  One would prefer perhaps the existence of shorter tails on average, 
however we are careful to note here that, in many cases for the XXZ model, the average retained loop length can be quite small (only several vertices).  It would clearly be interesting to 
address this ratio in other, more complicated Hamiltonians, which tend to produce larger loops.

Other characterizations of the short-loop are possible, in particular a comparison of the (retained) loop length
to the long-loop length.  In this case, several definitions of loop ``length'' are possible.  For example, we 
might be interested in the short-loop $\ell_{\rm sl}$ mentioned above, compared to a ``conventional''
definition of the long-loop length $\ell_{\rm ll}$ 
(which does not account for bounces or backtracking such as in 
Fig.~\ref{BounceP}).  Again, this comparison is expected to depend highly on the model, lattice size and dimension,
and parameter region one simulating.  We find in this simple XXZ model, for example, that the ratio
$\ell_{\rm ll}/\ell_{\rm sl} \approx 5$ to 10, for simulations with $h=0$ and $\Delta>1$ (depending on $L$ and
other factors).  For simulations with $\Delta=0$ and finite $h>0$, $\ell_{\rm ll}/\ell_{\rm sl}$ begins at approximately $20$ and
falls off as $h$ increases.  Because we choose an equilibriation condition that the number of vertices traversed be constant and equal (discussed more below), the ratio of the number of loops performed by the long-loop to the small-loop reflects very closely the inverse of the ratio $\ell_{\rm ll}/\ell_{\rm sl}$.  Depending on implementation, the proportional amount of CPU time spent on the short-loop algorithm can also be a significant constant multiple of this ratio.

We turn to what is perhaps the most quantitative indicator of simulation performance -- measurements of {\em autocorrelation} functions 
for observable parameters \cite{LiuMC}.  
The autocorrelation for a Monte Carlo time series of observables
${\mathcal O}(1)$, ${\mathcal O}(2)$,..., is defined with the normalized correlation function,
\begin{equation}
A[{\mathcal O}](t) = \frac{ \langle ({\mathcal O}(i+t){\mathcal O}(t) \rangle - \langle {\mathcal O}(i) \rangle^2 }{\langle {\mathcal O}(i)^2 \rangle},
\label{AutoC}
\end{equation}
where the averages are over the Monte Carlo ``time'' steps $i$ (elements of the Markov chain).  Higher 
autocorrelations imply that series elements are less independent, while small autocorrelation times 
are a necessary condition for a simulation to be ergodic in a specific region of configuration space. 

Before we proceed, we caution that one must be careful to note that quantitative values of autocorrelation 
functions are highly dependent on several simulation variables, which might be most concisely summed up as the
definition of a {\em Monte Carlo step} (MCS).  The MCS defines the increment $i$ in the definition 
Eq.~(\ref{AutoC}), and hence critically affects the measurement of this quantity.  In the SSE QMC, a MCS is
typically defined as a diagonal update (mentioned above), followed by a {\em number} of operator-loops.  Changes incurred within these updates are mapped onto the stored
basis state $| \alpha \rangle$ and operator string $S_M$, at which point the MCS is completed (and subsequently repeated).  The number of operator-loops is perhaps the most important variable in the definition of a MCS, and upon consideration
it immediately becomes clear that this number is potentially defined much differently in a long versus a 
short loop, since the loop-length discussed above varies considerably between the two.  
For example, a typical \cite{DIRloop} way to define a MCS is that it contains a number of loops ($N_{\rm loop}$) that on 
average will traverse each vertex-leg in the linked-list {\em once}; 
\begin{equation}
\bar{N}_{\rm leg} = c_{l} \cdot N_{\rm loop} \cdot \bar{\ell}_{\rm loop},
\label{NeqCond}
\end{equation}
i.e.~the constant $c_l$ is set to 1.  Here, $\bar{N}_{\rm leg}$ is the number of vertices ($n$)  in the expansion
multiplied by the number of legs per vertex (four for the simple XXZ model), and the average loop length (the number of legs visited by each loop: $\bar{\ell}_{\rm loop}$) may be approximated during equilibriation time.
A smaller $c_l$ will in general result in larger autocorrelations, since by definition each MCS traverses
less vertex legs, resulting in more dependence between configurations in adjacent QMC steps.  
In the following results we set $c_l=0.25$, 
and adjust $N_{\rm loop}$ during equilibriation to satisfy Eq.~(\ref{NeqCond}).
This value of $c_l$ is smaller than convention, however it increases our autocorrelations to a manageable value in this simple model.

Another consideration not taken into account by simple autocorrelation function comparisons is the CPU
effort involved when the definition of a MCS varies significantly, as in our case.  Clearly, 
the short loop algorithm involves both the overhead of storage of the stack data structure (containing both
the loop and the tail), as well as the additional computational effort
of erasing the tail at the end of loop construction.  
Indeed we observe that a short-loop MCS takes more CPU time than an analagously-defined long-loop MCS,
although the extent to which depends highly on algorithmic implementation and compiler optimization.
Nonetheless, we keep this in mind in the following discussion.

\begin{figure}[floatfix] 
\includegraphics[width=3.2 in]{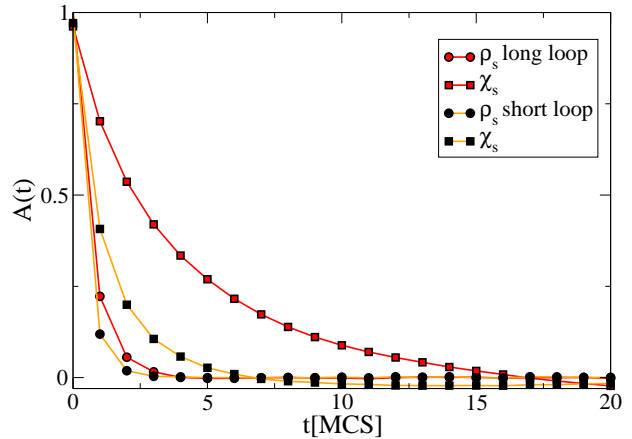}
\caption{(color online) 
Autocorrelations of $\rho_s$ and $\chi_s$ for the XXZ model with $L=16$, $\beta=32$, $h=0$ and $\Delta=1.1$ for both the long-loop (colored symbols) and short-loop (solid symbols) algorithms. Significant improvement of the autocorrelations for the short-loop  over that for the long-loop algorithm is observed. 
\label{AutoCF}}
\end{figure}

Figure~\ref{AutoCF} illustrates autocorrelations versus Monte Carlo time step for two common observables, 
the spin stiffness $\rho_s$ \cite{PolCep} and the staggered susceptibility $\chi_s$ \cite{sse1}, for the XXZ model with parameters $L=16$, $h=0$, and $\Delta=1.1$.
It is already apparent that the short-loop considerably improves autocorrelations for both
observables in this simple demonstration.  The CPU time involved in the short-loop run was larger 
than the long-loop run by a factor of 4 in this case.  Upon reflection however, it is perhaps remarkable that the short-loop gives any improvements to autocorrelations whatsoever, since presumably (as chosen by the equilibrium condition Eq.~(\ref{NeqCond})) the {\rm same} average number of vertex legs have been traversed by both algorithms.  This is possibly a measure of the degree to which the elimination of re-tracing (Figs.~\ref{SLloop1}), or the flipping of many {\it independent} short-loops (discussed previously), give efficiency improvements over the long-loop algorithm.  It would clearly be interesting to study this issue in more detail in the future.
\begin{figure}[floatfix] 
\includegraphics[width=3.0 in]{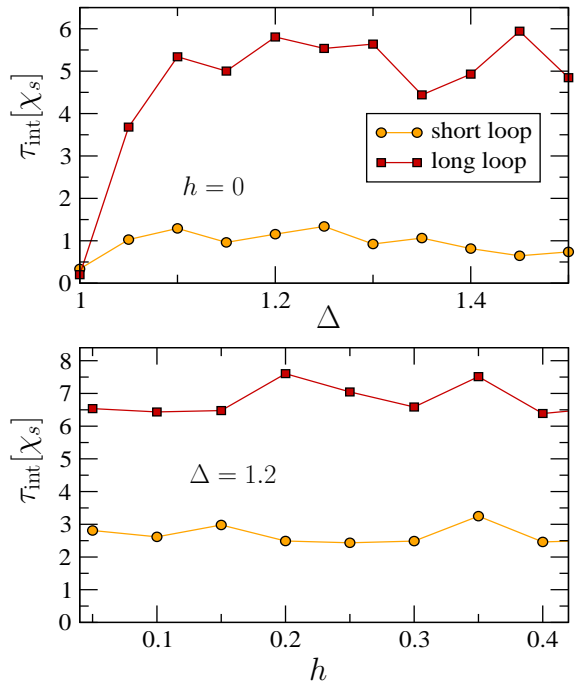}
\caption{(color online) 
Integrated autocorrelation time for a $L=16$ system at $\beta=32$.  Error bars, although not plotted
explicitly, are evident in the magnitude of fluctuations of the data points.
\label{DHtau}}
\end{figure}

\begin{figure} 
\includegraphics[width=3.0 in]{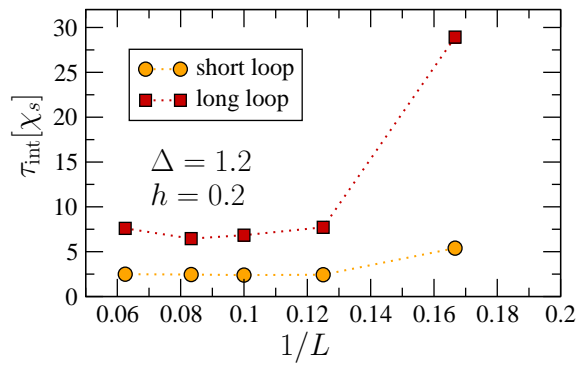}
\caption{(color online) 
Integrated autocorrelation time for simulations with $\Delta=1.2$ and $h=0.2$, versus the inverse linear
system size.  The inverse temperature of each run is set at $\beta=2L$.   Error bars are roughly equivalent to the symbol size.
\label{FSS_chi}}
\end{figure}

We concentrate now on autocorrelation measurements of the slowest decay in Fig.~\ref{AutoCF},
the staggered spin susceptibility.
To further characterize algorithmic effeciency, we look at integrated autocorrelation times, defined as
\begin{equation}
\tau_{\rm int} [\mathcal{O}] = \frac{1}{2} + \sum_{t=1}^{\infty} A[\mathcal{O}](t),
\label{TAU}
\end{equation}
using $\mathcal{O} = \chi_s$. 
Figure~\ref{DHtau} illustrates this calculation for short- versus long-loops as one sweeps in parameters 
$\Delta$ and $h$, for $L=16$, using one particular solution to the directed loop equations which minimize bounces \cite{DIRloop}.  
As evident from Fig.~\ref{FSS_chi}, this system size is large enough that any remaining finite-size effects
are obscured by statistical errors (note also that the computational effort of the SSE QMC scales linearly in both $\beta$ and $L$).
The quantitative value
of $\tau_{\rm int}[\chi_s]$ depends highly on the definition of the MCS ($c_l$), as well as the directed
loop equations (using a heat-bath solution \cite{DIRloop} increases $\tau_{\rm int}[\chi_s]$ considerably).
In some cases, for example in certain bounce-free regions of parameter space, integrated 
autocorrelations are very small and the difference between short- and long-loop performance is
lost in the statistical errors.  However, for another large region of phase space (such as that illustrated
in Fig.~\ref{DHtau}), the general trend is that the short-loop algorithm outperforms the long-loops
in terms of $\tau_{\rm int}[\chi_s]$.  Surprisingly, even in the simple XXZ model, we were unable to find regions of parameter space where $\tau_{\rm int}[\chi_s]$ is significantly larger for the short-loop algorithm than for the long-loop algorithm.  That being said, we observe that in our implementations
of the short-loop algorithm, the amount of CPU time required to 
produce results such as in Fig.~\ref{DHtau} are significantly larger for the short-loop code: typically
by a factor of four or more.  Memory allocation is also larger for the short-loops, although as with most
QMC simulations still relatively small compared to hardware constraints, offering no real disadvantage
over the conventional long-loop algorithm.

\begin{figure}[bht] 
\includegraphics[width=3.0 in]{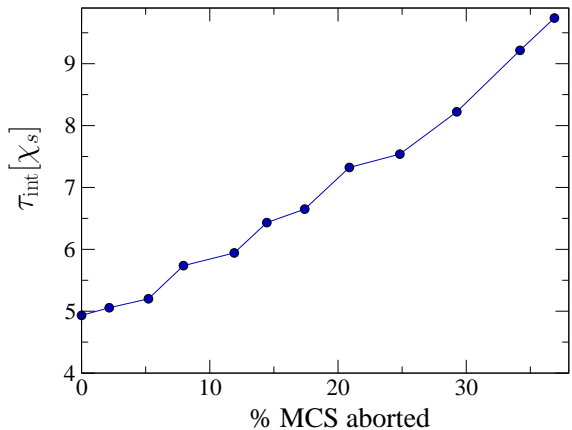}
\caption{(color online) 
The integrated autocorrelation time for a $L=8$, $\beta=16$ system with $h=0$ and $\Delta=1.4$.  The 
x-axis is the percentage of Monte Carlo steps that are aborted due to the termination of a long-loop.
This truncation percentage is controlled by restricting the maximum loop length with Eq.~(\ref{Llength}).
\label{TruncT}}
\end{figure}

Recall, the purpose
of the short-loop is perhaps not to bring significant  efficiency gains to all models, rather to those models where  long-loop length tends to get excessive, causing in some cases the loop tend to be truncated.
Figure~\ref{TruncT} demonstrates the fact that this practice of truncating long-loops in the SSE QMC 
results in a systematic increase in the integrated autocorrelation time.  By comparison, the same set
of parameters using the short-loop algorithm results in a $\tau_{\rm int}[\chi_s] \approx 1$,
when run using the same condition ($c_l=0.25$ in Eq.~(\ref{NeqCond})) to define the QMC (with a smaller number of short-loops, this value will increase).  Thus, it is clear that the advantage in using the short-loop algorithm will only increase
in events where long-loops are observed to become aborted (or excessively long), such as those expected to occur on more complicated quantum models.

\section{Discussion}

Motivated by the success of short-loop algorithms in Monte Carlo simulations of classical vertex and ice models, we have presented an adaptation of the short-loop algorithm for use in QMC simulations of quantum lattice Hamiltonians.  
This quantum short-loop algorithm is a modification of the conventional (``long'') quantum loop or worm algorithm, whereby smaller clusters in the $d+1$-dimensional simulation cell may be constructed and 
updated.  Short-loops are defined by a modified construction algorithm, where a propagating defect
closes a loop upon crossing a part of its previously-visited path.  Unlike the conventional long-loop, this results in a {\em dangling tail} that must be removed before the QMC algorithm can
continue.
Within the SSE QMC framework, we introduced the general algorithmic rules and data structures required for constructing and updating short loops (including an additional stack to store the loop under construction), and outline some expected advantages and disadvantages of the new algorithm as compared to the conventional long-loop algorithm. 

Using a C++ implementation of the short-loop algorithm in the simple square-lattice $S=1/2$ XXZ model, we characterized key aspects of simulation performance, and compare to the conventional long-loop algorithm using identical parameters.
Remarkably, the short-loop algorithm is observed to give much smaller
autocorrelation times - a key indicator that this modification results in an
increase in simulation efficiency.  However, with this improvement in
autocorrelation time comes a significant increase in CPU effort (and to a
lesser extent memory usage).  Hence, before using the short-loop algorithm
in large-scale QMC simulations, one must be careful in identifying models
and parameter regions where this compromise becomes favorable.

Significant work still remains to be done in optimizing and characterizing the short-loop algorithm, particularly in different QMC flavors and on more complicated quantum models.  Although conventional estimators like those discussed here will remain unbiased by the short-loop algorithm, it remains to be determined whether certain schemes to measure Green's functions and dynamical properties are affected by the smaller loops that are generated \cite{TroyerGF}.  
In the immediate future, the quantum short-loop algorithm will likely be most useful in specific complicated models (e.g.~those with long-range interactions in the Hamiltonian) where conventional long-loops are observed to behave poorly, rather than as a means of improving efficiency in the general case.
Eventually, a more wide-spread adoption of the short-loop may be warranted, although further implementation and characterization on additional models will be required to more clearly identify its strengths and weaknesses. 

\begin{acknowledgments}
We thank A. Sandvik, M. Troyer and S. Wessel for valuable discussions. YJK thanks the hospitality of KITP, Santa Barbara, where part of this work was done, and National Center for High Performance Computing of Taiwan, and Computer and Information Networking Center at NTU for the support of high-performance computing facilities. 
RGM acknowledges use of the facilities of the Shared Hierarchical Academic Research Computing Network (SHARCNET).
This work was partly supported by NSC of Taiwan under grant No. 96-2112-M-002-010, NSF  under grant No. PHY05-51164 (YJK), and
the U.S. Department of Energy, contract DE-AC05-00OR22725 with ORNL,
managed by UT-Battelle, LLC (RGM).\end{acknowledgments}

\bibliography{rmBiblio}
  
\end{document}